\documentclass[aps, prd, 10pt, onecolumn, floatfix, altaffilletter, superscriptaddress, preprintnumbers, tightenlines, showpacs, showkeys, nofootinbib, notitlepage]{revtex4-2}

\usepackage[a4paper, centering, hmargin=2.5cm, vmargin=2.5cm]{geometry} 

\usepackage[normalem]{ulem} 
\usepackage{amsmath, amssymb, amsthm} 
\usepackage{mathtools} 
\usepackage{graphicx} 
\usepackage[x11names, table]{xcolor} 
\graphicspath{{./figs/}} 
\usepackage{booktabs} 
\usepackage{siunitx} 
\usepackage{array} 
\usepackage{threeparttable} 
\usepackage{aecompl} 
\usepackage{appendix} 
\usepackage[colorlinks=true,citecolor=blue,linkcolor=blue]{hyperref} 
\usepackage{orcidlink} 
\usepackage{mathrsfs} 
\usepackage{slashed} 
\usepackage{multirow} 
\usepackage{enumitem} 
\usepackage{cancel} 
\usepackage{tikz} 
\usepackage[spanish, english]{babel} 
\usepackage{textgreek} 
\usepackage{bm} 
\usepackage[capitalise, english]{cleveref} 
\usepackage{hhline} 
\usepackage{soul} 
\usepackage{xspace} 
\usepackage{scalerel} 
\usepackage{bbding} 
\usepackage{adjustbox} 
\usepackage{upgreek} 

\makeatletter

    \def\CT@@do@color{%
      \global\let\CT@do@color\relax
            \@tempdima\wd\z@
            \advance\@tempdima\@tempdimb
            \advance\@tempdima\@tempdimc
    \advance\@tempdimb\tabcolsep
    \advance\@tempdimc\tabcolsep
    \advance\@tempdima2\tabcolsep
            \kern-\@tempdimb
            \leaders\vrule
                    \hskip\@tempdima\@plus  1fill
            \kern-\@tempdimc
            \hskip-\wd\z@ \@plus -1fill }
    \makeatother

\DeclareSIUnit\parsec{pc}

\hypersetup{colorlinks,citecolor= nicered,linkcolor= blue}
\definecolor{nicered}{rgb}{0.7,0.1,0.1}
\definecolor{nicegreen}{rgb}{0.1,0.5,0.1}

\def\cevns{CE$\upnu$NS}

\newcommand{\GV}{$\mathcal{G}_V$}

\definecolor{myblue}{cmyk}{0.65, 0.37, 0.0, 0.19}
\definecolor{myred}{cmyk}{0.00, 1.00, 1.0, 0.46}
\definecolor{mymagenta}{cmyk}{0, 1, 0, 0.1}
\def\AT#1{{\color{mymagenta}#1}}

\AtBeginDocument{\hypersetup{citecolor=myblue,linkcolor=myblue,urlcolor=myblue}}

\begin{document}

\newcommand{\AddrIFIC}{%
Instituto de F\'{i}sica Corpuscular (CSIC-Universitat de Val\`{e}ncia), Parc Cient\'ific UV C/ Catedr\'atico Jos\'e Beltr\'an, 2, E-46980 Paterna (Valencia), Spain
}
\newcommand{\AddrDepFUV}{%
Departament de F\'isica Teòrica, Universitat de Val\`{e}ncia, 46100 Burjassot, Spain}

\newcommand{\AddrKolk}{%
Ramakrishna Mission Residential College (Autonomous) \& Vivekananda Centre for Research, Narendrapur, Kolkata 700103, India}
             
\newcommand{\AddrIacs}{%
School of Physical Sciences, Indian Association for the Cultivation of Science,
2A \& 2B Raja S.C Mullick Road, Kolkata-700032, India}
\title{Neutrino fog in the light dark sector: the role of isospin violation}

\author{V\'ictor Mart\'in Lozano~\orcidlink{0000-0002-9601-0347}}\email{victor.lozano@ific.uv.es}
\affiliation{\AddrIFIC}
\affiliation{\AddrDepFUV}

\author{Shankar Pramanik~\orcidlink{0000-0000-0000-0000}}\email{shankar1103pramanik@gmail.com}
\affiliation{\AddrKolk}

\author{Soumya Sadhukhan~\orcidlink{0000-0002-2958-050X}}\email{soumya.sadhukhan@rkmrc.in}
\affiliation{\AddrKolk}
\affiliation{\AddrIacs}

\author{Adri\'an Terrones~\orcidlink{0000-0003-1636-3335}}\email{adrian.terrones@ific.uv.es}
\affiliation{\AddrIFIC}


\begin{abstract}
\noindent

Dark matter (DM) direct detection is now standing at an interesting juncture, where the Standard Model (SM) neutrino background and the upper bound on the DM signal cross section are starting to overlap in the DM mass region around 10 GeV. The neutrino floor, which defines the extent of the neutrino background, can be modified in different Beyond Standard Model (BSM) setups. We work in a simple BSM dark sector extension of the SM visible sector, where isospin-violating interactions occur naturally. In this model, both DM and neutrinos have, in general, isospin-violating (IV) interactions with nuclei, through a newly added local U(1) gauge boson $Z^{\prime}$. Depending on the choice of the model parameters, the coherent elastic neutrino-nucleus scattering (CE$\upnu$NS) cross section can either increase or decrease, shifting up or down the neutrino floor in the parameter space. The same is true for the DM experimental upper bound, whose change is driven exclusively by the IV parameter $f_n/f_p$. Several scenarios are constructed, based on the interplay between the two regions and the allowed parameter space left between them, and discussed. The potential observation of solar neutrinos in DM direct detection experiments is also discussed in the context of our framework.
\end{abstract}
\maketitle
\section{Introduction}
\label{sec:intro}
Even if dark matter (DM) has been established to form 85\% of the matter density of the Universe, we are yet to observe it in DM direct detection experiments. In these, we expect DM particles around us to scatter off the detector material, inducing nuclear recoils within the detector. Great efforts have been made in this direction, culminating in the ton-scale generation of xenon (Xe) experiments such as LZ~\cite{LZ:2022lsv}, PandaX~\cite{PandaX-4T:2021bab,PandaX:2024qfu} and XENONnT~\cite{XENON:2025vwd}, imposing stringent bounds to the DM parameter space.

At the same time, neutrinos, predominantly those produced in the Sun, can also scatter off the detector material's nuclei through neutral current processes, known as coherent elastic neutrino-nucleus scattering (\cevns). These neutrino-induced recoils could potentially act as background for prospective DM signals.
In this context, the neutrino floor or neutrino fog~\cite{OHare:2021utq} is defined as the lower boundary of the DM-nucleon scattering cross section, as a function of DM mass, above which it is possible to definitely identify a recoil as a DM signal. Below that boundary, neutrinos become a background, making DM searches more challenging, while at the same time allowing for the exploration of new physics in the neutrino sector~\cite{AristizabalSierra:2019ykk, AristizabalSierra:2024nwf, DeRomeri:2024iaw,DeRomeri:2024hvc,Blanco-Mas:2024ale}. 
Recent searches by DM direct detection experiments have set bounds on the DM-nucleon scattering cross section that are becoming closer to the neutrino floor~\cite{LZ:2022lsv,XENON:2025vwd,PandaX:2024qfu,PandaX-4T:2021bab}. Furthermore, low-energy threshold searches by PandaX~\cite{PandaX:2022xqx,PandaX:2024muv} and XENONnT~\cite{XENON:2020gfr,XENON:2024ijk} found indications of solar neutrino events, albeit with very low significant statistics ($\sim 2.7~\sigma$). 

This is leading us to different but connected conundrums: where are the Standard Model (SM) neutrino interactions that were supposed to be present even when no DM signal is present?  Has the background-free DM parameter space totally disappeared? Are the DM and the neutrino sectors connected?

In the study of DM scenarios in both theoretical and experimental contexts, the assumption that DM interacts equally with protons and neutrons, known as isospin-conserving interactions, is often made. However, this might not be the case for the new physics of DM interactions~\cite{Giuliani:2005my,Feng:2011vu}. A clear example of isospin-violating (IV) interactions can be found in the SM itself, where the $Z$ boson interactions to neutrons are stronger than to protons. Isospin-violating interactions can arise naturally in numerous theoretical scenarios of DM sectors~\cite{Chun:2010ve,Frandsen:2011cg,Gao:2011ka,He:2011gc,He:2013suk,Feng:2013vod,Belanger:2013tla,Hamaguchi:2014pja,Chen:2014tka,Lozano:2015vlv,Crivellin:2015bva,Drozd:2015gda,Chang:2017gla,Martin-Lozano:2018sxw,Li:2019sty,Lozano:2021zbu}. One of the most visible effects that distinguishes isospin-violating models from isospin-conserving ones is the need to reinterpret DM direct detection bounds, since they are typically derived under the assumption of isospin-conserving interactions \cite{Tovey:2000mm,Zheng:2014nga,Yaguna:2016bga,Kelso:2017gib,Yaguna:2019llp,Cheek:2023zhv,Kumar:2023lrj}. The presence of isospin-violating interactions can either relax or make the bounds more stringent. One of the most studied isospin-violating scenarios is one that connects the DM sector to the SM through a $Z'$ coming from a $\mathrm{U}(1)$ gauge extension. In this case, the $Z'$ couplings to up and down quarks can be different and, as a consequence, the DM particles interact differently with protons and neutrons. On the other hand, the presence of a $\mathrm{U}(1)$ extension can also affect  SM particles by introducing new interactions. In particular, neutrino interactions with quarks will be affected, and thus their scattering off nuclei. If neutrino interactions with nuclei get modified, then the neutrino floor region in the parameter space will be placed differently compared to the one predicted by SM interactions alone.

In this work, we will study the phenomenology of a generic $Z'$ isospin-violating model that connects the DM sector to the SM. In particular, we will study how different amounts of isospin violation can relax or enhance direct detection bounds. Furthermore, we will also study the impact of such models on the neutrino sector and consequently on the neutrino floor. Taking into account both physical observables, we will study how the remaining background-free scenario for DM is affected by isospin-violating interactions.


This paper is organized as follows. In Sec.~\ref{sec:model} we introduce the isospin-violating DM framework that we will study. Then, the phenomenology of isospin-violating DM models in light of direct detection experiments is discussed in Sec.~\ref{sec:DMD}. We present our results in Sec.~\ref{sec:results} for different scenarios, and we draw our conclusions in Sec.~\ref{sec:conclusions}.



\section{Isospin-Violating Dark Matter Framework}
\label{sec:model}

In order to study the phenomenology of isospin-violating dark matter interactions, we extend the SM with a simplified model that adds a fermionic DM candidate, $\chi$, with mass $m_\chi$; and a spin-1 mediator, $Z'$, with mass $m_{Z'}$. The general Lagrangian term considered here can be read as
\begin{equation}
    \mathcal{L}\supset Z'_\mu \sum_f \bar{f}\gamma^\mu(g_V^f+g_A^f\gamma_5)f + Z'_\mu\bar{\chi}\gamma^\mu(g_V^\chi+g_A^\chi\gamma_5)\chi  + \frac{1}{2}m_{Z'}^2Z'_\mu Z'^\mu + m_\chi \bar{\chi}\chi \,,
    \label{eq:lagr}
\end{equation}
where $g_{V,A}$ are vector and axial-vector couplings, respectively, and $f$ runs over the SM fermions. The choice of this Lagrangian is motivated by the simplified DM models used in colliders~\cite{Abdallah:2015ter,DeSimone:2016fbz}, although we are aware that it could develop problems in the ultraviolet (UV) regime~\cite{Kahlhoefer:2015bea,DEramo:2016gos,Ismail:2016tod,Ellis:2017tkh}. Nonetheless, Eq.~\eqref{eq:lagr} can be considered an effective Lagrangian with a UV completion at some high scale, for example, with a St\"uckelberg mass realization for the $Z'$ (see, for instance, Ref.~\cite{Lozano:2015vlv}). In the following, we set $g_A^\chi=0$ to avoid any problems with unitarity~\cite{Kahlhoefer:2015bea}. This choice, $g_A^\chi=0$, avoids spin-dependent (SD) interactions, so we assume only spin-independent (SI) processes take place in the direct detection experiments.\footnote{If $g_A^\chi\neq 0$ is chosen, the number of events coming from SD processes also has to be considered when computing the total number of DM events hitting the detector.}

In general, the couplings of the $Z'$ to the up- and down-type quarks can be arbitrarily different, inducing different couplings to protons and neutrons, thus leading to what we refer to as isospin violation. The effective SI couplings of DM to protons and neutrons, respectively, can be written as~\cite{Chun:2010ve,Lozano:2015vlv}
\begin{equation}
f_{p,n} = g^\chi_V\frac{(2g_V^{u,d} +g_V^{d,u} )}{2m^2_{Z'}} \,.
\end{equation}
We can then define the amount of isospin violation as the ratio between the neutron and proton effective couplings, which in this model reads
\begin{equation}
    \frac{f_n}{f_p} = \frac{ 2g_V^{d} +g_V^{u} }{ 2g_V^{u} +g_V^{d} } \,.
    \label{eq:fnfp}
\end{equation}
The interaction is isospin-conserving ($f_n/f_p = 1$) if the couplings to the up and down quarks are equal, $g_V^u=g_V^d$, but in general a different choice of up and down quark couplings will break the isospin symmetry.

This particular kind of model can be subject to different bounds, such as from collider experiments. For example, in the low mass range ($m_{Z'}\lesssim 200$ GeV), the most relevant constraints are those coming from LEP~\cite{DELPHI:1994ufk}, and for very low masses  LHCb~\cite{LHCb:2017trq} and BaBar~\cite{BaBar:2014zli} set very stringent bounds to the new sector couplings. In the high mass range ($m_{Z'}\gtrsim 200$ GeV) there are several searches from both ATLAS and CMS in different final state channels, the most competitive being those involving leptons in the final state. Regarding the neutrino sector, its couplings to the $Z'$ can induce new observables that can be tested at both low and high energies~\cite{DeRomeri:2022twg,Lozano:2025ekx}.

\section{Isospin-Violating Dark Matter at Direct Detection Experiments}
\label{sec:DMD}

Direct detection experiments are sensitive to particles that can interact with nuclei. In particular, given the isolation of these experiments, particles that interact weakly with matter can be observed by their detectors, which is the case for DM particles and neutrinos. The effective Lagrangian of Eq.~\eqref{eq:lagr} introduces interactions of the new mediator with the components of the nucleus, the DM sector, and neutrinos, coupling DM to ordinary matter and modifying the SM neutrino-nucleus interactions. All these new interactions can be tested in direct detection experiments. In this section, we will describe such interactions and how they affect the experiment's observables through the scattering with nuclei, modifying both the DM direct detection exclusion bounds and the neutrino fog in those experiments.

\subsection{Dark Matter Exclusion Bounds}
The event rate of a dark matter particle interacting with a given detector composed of a nucleus with $Z$ protons, $A$ nucleons, and $A-Z$ neutrons can be written as
\begin{equation}
    R= \sigma^0_A \times I_A\,,
    \label{eq:rate}
\end{equation}
where $\sigma^0_A$ is the spin independent dark matter-nucleus cross section at zero momentum transfer and $I_A$ is a function that depends on the astrophysical, nuclear and experimental physics. Assuming that our DM candidate can interact differently with neutrons and protons, one may take into account the natural abundances of the isotopes of each detector composition. Taking that into account, Eq.~\eqref{eq:rate} reads,
\begin{equation}
    R= \sum_i \eta_i\sigma_{A_i}^0 \times I_{A_i}\,,
    \label{eq:rateabun}
\end{equation}
where $i$ indexes the isotopes of the detector material, and $\eta_i$ is the natural abundance of the isotope $A_i$. The SI cross section at zero momentum transfer for a given $A_i$ is defined as
\begin{equation}
    \sigma^0_{A_i}=\frac{4\mu^2_{A_i}}{\pi}[f_p Z + f_n (A_i-Z)]^2\,,
    \label{eq:sigma0}
\end{equation}
where $f_{p,n}$ are the effective couplings of the DM particle to protons and neutrons, respectively, and $\mu_{A_i}=m_{A_i}m_\chi/(m_{A_i}+m_\chi)$ is the reduced mass of the DM-nucleus system. One can see that for a specific isotope the cross section vanishes if $f_n/f_p=-Z/(A_i-Z)$. If this happens, the total event rate of dark matter is driven to zero for that isotope. However, in a realistic detector composed of a specific material, the different abundances of the isotopes prevent the cross section from vanishing, even if the total event rate remains small.

\begin{figure*}[ht]
	\centering
        \includegraphics[width=.44\textwidth]{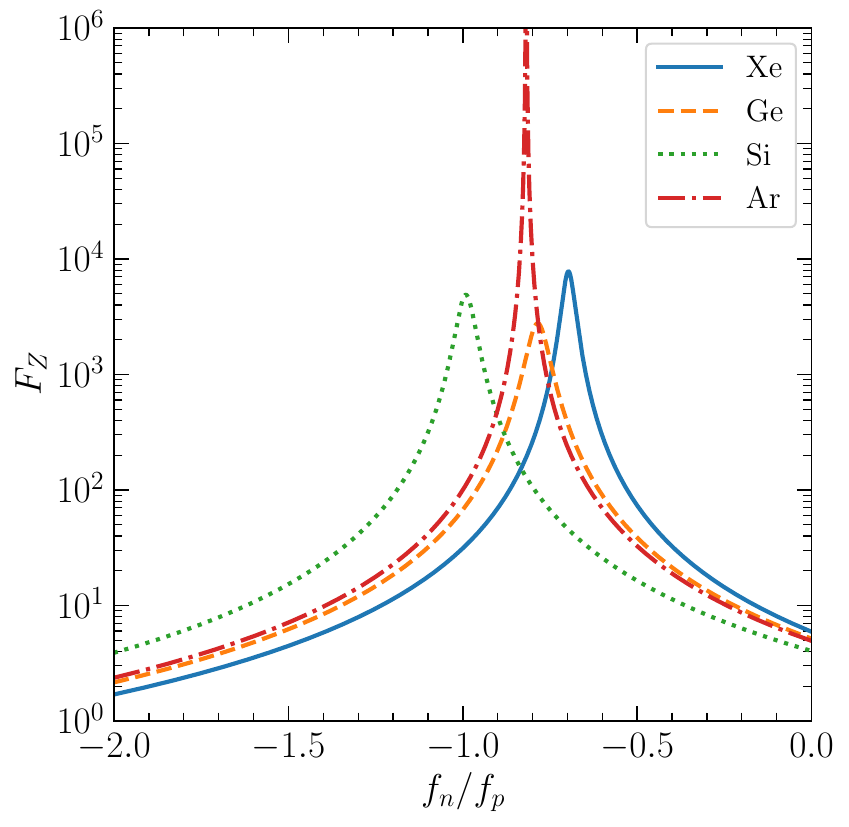}\hspace{1.0cm}
        \includegraphics[width=.44\textwidth]{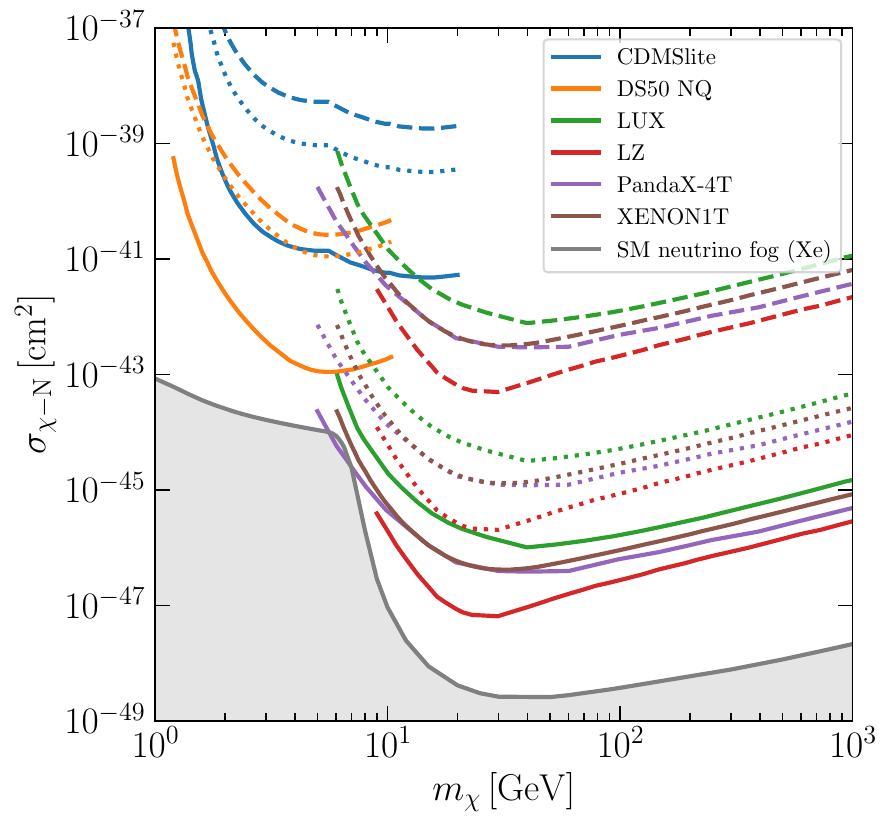}
	\caption{\textbf{Left Panel:} $F_Z$ factor as a function of the isospin violation parameter $f_n/f_p$ for Xe (solid blue), Ge (dashed orange), Si (dotted green) and Ar (dash-dotted red). A DM mass $m_\chi=10$ GeV was chosen to produce the figure, but note that variations of $m_\chi$ in the GeV--TeV range produce negligible changes in $F_Z$. \textbf{Right Panel:} DM-nucleon spin independent cross section as a function of DM mass. Different colors correspond to different experiments: CDMSlite (blue), DarkSide50 (orange), LUX (green), LZ (red), PandaX-4T (purple) and Xenon1T (brown). The different line styles correspond to different values of the isospin violation parameter: $f_n/f_p=1$ (solid), $f_n/f_p=-0.7$ (dashed), and $f_n/f_p=-1$ (dotted). The SM neutrino fog for Xe detectors is shown in gray.}
	\label{fig:dz}
\end{figure*}

Typically, experimental exclusion limits are presented in terms of the DM-nucleon cross section, $\sigma_N^Z$, assuming no isospin violation in the DM sector ($f_p=f_n$), and therefore $\sigma_N^Z=\sigma_p$, where $\sigma_p$ is the spin independent DM-proton cross section defined as $\sigma_p=4\mu_p f_p/\pi$.
However, in scenarios with isospin violation, the spin independent DM-nucleus cross section would have the form
\begin{equation}
\sigma_N^Z=\sigma_p \frac{\sum_i \eta_i \mu^2_{A_i}[Z+(A_i-Z)f_n/f_p]^2}{\sum_i \eta_i\mu^2_{A_i}A_i^2}\,,
\end{equation}
where the relative abundances $\eta_i$ of the isotopes of the element are taken into account. Then, we can conveniently define the $F_Z$ factor as the following ratio,
\begin{equation}
    F_Z\equiv \frac{\sigma_p}{\sigma_N^Z}=\frac{\sum_i \eta_i\mu^2_{A_i}A_i^2}{\sum_i \eta_i \mu^2_{A_i}[Z+(A_i-Z)f_n/f_p]^2}\,.
    \label{eq:DZ}
\end{equation}
The meaning of this ratio is the factor by which the sensitivity of direct detection experiment results is suppressed in the presence of IV. As we can see, this factor depends only on the number of protons of the element, $Z$, the number of nucleons, $A_i$, and the relative abundances of the isotopes of the element, $\eta_i$, being totally insensitive to the DM and mediator masses, and its couplings. In Table~\ref{tab:abundances} we can find the isotope abundances for xenon, germanium, silicon and argon. In the left panel of Fig.~\ref{fig:dz} we can see the $F_Z$ factor for these four elements. For these elements, we have a maximal suppression in the event rate when the isospin violation parameter falls in the region $f_n/f_p \in (-1.0,-0.7)$. One may notice that $F_Z$ is not driven to zero for the majority of the elements due to the different abundances of isotopes, with the exception of argon.

\begin{table*}[ht]
\begin{center}

  \renewcommand{\arraystretch}{1}
  \begin{adjustbox}{max width=\textwidth}
  \begin{tabular}{c | c  c c c c c c c }
    \hline
\multirow{2}{*}{${}_{54}$Xe} & \multicolumn{1}{l}{$A_i$} & \multicolumn{1}{l}{128} & \multicolumn{1}{l}{129} & \multicolumn{1}{l}{130} & \multicolumn{1}{l}{131} & \multicolumn{1}{l}{132} & \multicolumn{1}{l}{134} & \multicolumn{1}{l}{136} \\\cline{2-9}
                     & \multicolumn{1}{l}{$\eta_i$} & \multicolumn{1}{l}{1.9} & \multicolumn{1}{l}{26} & \multicolumn{1}{l}{4.1} & \multicolumn{1}{l}{21} & \multicolumn{1}{l}{27} & \multicolumn{1}{l}{10} & \multicolumn{1}{l}{8.9} \\\cline{1-9}
\multirow{2}{*}{${}_{32}$Ge} & \multicolumn{1}{l}{$A_i$} & \multicolumn{1}{l}{70} & \multicolumn{1}{l}{72} & \multicolumn{1}{l}{73} & \multicolumn{1}{l}{74} & \multicolumn{1}{l}{76} & \multicolumn{1}{l}{} & \multicolumn{1}{l}{} \\\cline{2-7}
                     & \multicolumn{1}{l}{$\eta_i$} & \multicolumn{1}{l}{21} & \multicolumn{1}{l}{28} & \multicolumn{1}{l}{7.7} & \multicolumn{1}{l}{36} & \multicolumn{1}{l}{7.4} & \multicolumn{1}{l}{} & \multicolumn{1}{l}{}
                                 \\\cline{1-7}
\multirow{2}{*}{${}_{28}$Si} & \multicolumn{1}{l}{$A_i$} & \multicolumn{1}{l}{28} & \multicolumn{1}{l}{29} & \multicolumn{1}{l}{30} & \multicolumn{1}{l}{} & \multicolumn{1}{l}{} & \multicolumn{1}{l}{} & \multicolumn{1}{l}{} \\\cline{2-5}
                    & \multicolumn{1}{l}{$\eta_i$} & \multicolumn{1}{l}{92} & \multicolumn{1}{l}{4.7} & \multicolumn{1}{l}{3.1} & \multicolumn{1}{l}{} & \multicolumn{1}{l}{} & \multicolumn{1}{l}{} & \multicolumn{1}{l}{}
                                 \\\cline{1-5}
\multirow{2}{*}{${}_{18}$Ar} & \multicolumn{1}{l}{$A_i$} & \multicolumn{1}{l}{40} & \multicolumn{1}{l}{} & \multicolumn{1}{l}{} & \multicolumn{1}{l}{} & \multicolumn{1}{l}{} & \multicolumn{1}{l}{} & \multicolumn{1}{l}{} \\\cline{2-3}
                    & \multicolumn{1}{l}{$\eta_i$} & \multicolumn{1}{l}{96} & \multicolumn{1}{l}{} & \multicolumn{1}{l}{} & \multicolumn{1}{l}{} & \multicolumn{1}{l}{} & \multicolumn{1}{l}{} & \multicolumn{1}{l}{}
                                 \\\cline{1-3}
  \end{tabular}
  \end{adjustbox}
  \caption{Abundances of each isotope of Xenon, Germanium, Silicon and Argon. Only abundances greater than 1\% are shown \cite{Feng:2011vu}.}
 \label{tab:abundances}
 \end{center}
\end{table*}

An example of how isospin-violating interactions may affect direct detection experimental results is shown in the right panel of Fig.~\ref{fig:dz}, where the different direct detection bounds of the spin independent DM-nucleon cross section are depicted as a function of the DM particle mass. The different colors of the lines represent the different experiments, namely CDMSlite (blue), DarkSide50 (orange), LUX (green), LZ (red), PandaX-4T (purple), and Xenon1T (brown), whereas the different line styles correspond to different isospin violating scenarios; solid lines correspond to the isospin conserving scenario, $f_n/f_p=1$, while dashed lines are for $f_n/f_p=-0.7$ (xenon-phobic scenario) and dotted for $f_n/f_p=-1$ (silicon-phobic scenario). We can see that, when the interactions do not conserve isospin, the bounds imposed by the experiments are in general weaker. For example, in the case where $f_n/f_p=-0.7$ (dashed) the DM-xenon cross section is the most affected by the degradation factor. The detectors based on xenon lose their sensitivity by almost four orders of magnitude, while those that are non-xenon based, such as CDMSlite and DarkSide50, are reduced around 2 orders of magnitude. If we set the isospin violation to $f_n/f_p=-1$, we can see that the reduction is less aggressive for all detectors. Nonetheless, the direct detection bounds are still less stringent than in the isospin-conserving case.


\subsection{Neutrino Fog}

\subsubsection{\texorpdfstring{\cevns~Cross Section}{CEνNS Cross Section}}

Due to their weak interactions with matter, neutrinos can easily pass through the shielding of direct detection experiments and interact with the nuclei of their detectors. These interactions produce a nuclear recoil that is difficult to differentiate from DM interactions. This process, mediated by neutral currents, is known as Coherent Elastic Neutrino-Nucleus Scattering (\cevns), and provides a useful tool to probe the effects of new physics.

In the limit where $m_N E_R \ll m_Z^2$ and $E_\nu \ll m_N$, the SM neutrino-nucleus scattering differential cross section is given by~\cite{Billard:2013qya}
\begin{equation}
    \dfrac{d\sigma^\nu_{\rm SM}}{dE_R}= \frac{G_F^2 m_N}{4\pi}\mathcal{F}(E_R)^2\left( Q_V^{\rm SM} \right) ^2\left(1-\frac{E_R m_N}{2E_\nu^2}\right)\,,
\end{equation}
where $G_F$ is the Fermi constant, $m_N$ is the nucleus mass, $Q_V^{\rm SM}=N-Z(1-4\sin^2\theta_w)$ is SM weak charge to nucleons, $E_R$ is the nucleus recoil energy, $E_\nu$ is the incident neutrino energy and $\mathcal{F}$ is the Helm nuclear form factor~\cite{Helm:1956zz, Lewin:1995rx}, given by
\begin{equation}
    \mathcal{F}(E_R)=3 \frac{j_1 \left( q(E_R)r_n \right)}{q(E_R)r_n} e^{-\frac{1}{2} s^2 q(E_R)^2},
\end{equation}
with $q(E_R) = \sqrt{2 m_N E_R}$ the momentum transfer, $s \simeq \SI{0.9}{\femto\meter}$ the nuclear skin thickness \cite{Lewin:1995rx}, $j_1 (x) = (\sin x - x \cos x)/x^2$ the first order spherical Bessel function of the first kind, and $r_n$ the effective nuclear radius, for which we assumed the linear parametrization $r_n = a_n A^{1/3} + b_n$, with $a_n = \SI{1.14}{\femto\meter}$ and $b_n = 0$.

In the presence of new physics that modifies the neutral currents, the neutrino-nucleus interaction will be affected. In particular, the differential cross section of the neutrino-nucleus interaction in the presence of a new vector boson can be written as~\cite{Cerdeno:2016sfi,Bertuzzo:2017tuf}
\begin{equation}
\frac{d\sigma_V^\nu}{dE_R}= \mathcal{G}_V\frac{d\sigma_{SM}^\nu}{dE_R}\,,
\label{eq:xi}
\end{equation}
where the factor $\mathcal{G}_V$, which indicates how the SM cross section gets modified, is given by\footnote{Note that our result matches with the one in Ref.~\cite{Cerdeno:2016sfi}, but does not agree in the sign of the cross term in version 3 of Ref.~\cite{Bertuzzo:2017tuf} in the arXiv.}
\begin{equation}
    \mathcal{G}_V = 1 - \frac{2\sqrt{2}}{G_F} \frac{Q_V}{Q_V^{SM}} \frac{g^{\nu}_V-g^{\nu}_A}{2m_NE_R+m_{Z'}^2} 
    + \left( \frac{2}{G_F} \right)^2 \left( \frac{Q_V}{Q_V^{SM}} \right)^2 \frac{(g^{\nu}_V)^2 + (g^{\nu}_A)^2}{ \left( 2m_NE_R+m_{Z'}^2 \right)^2 } \,.
    \label{eq:GV}
\end{equation}
$Q_V$ is the effective SI coupling of the new vector boson to a nucleus with $Z$ protons and $A-Z$ neutrons \cite{Bertuzzo:2017tuf},
\begin{equation}
    Q_V = Z \left( 2g_V^u + g_V^d \right) + (A-Z) \left( 2g_V^d + g_V^u \right) \,,
\end{equation}
where $g_V^{u,d}$ are the vector couplings of the up and down quarks to the $Z'$.

The modification of the factor \GV\ with respect to 1 will impact the total neutrino events that interact with nuclei. If  $\mathcal{G}_V<1$, the differential cross section will be smaller than the SM one, which will imply a reduction of the neutrino-nucleus interaction rate and hence a reduction in the number of neutrino events, due to the interference between the SM $Z$ boson and the new $Z'$ mediator. On the contrary, when $\mathcal{G}_V<1$ there is constructive interference, and therefore the differential cross section and the number of neutrino events will be greater than in the SM. 

\begin{figure*}[ht]
	\centering
        \includegraphics[width=.48\textwidth]{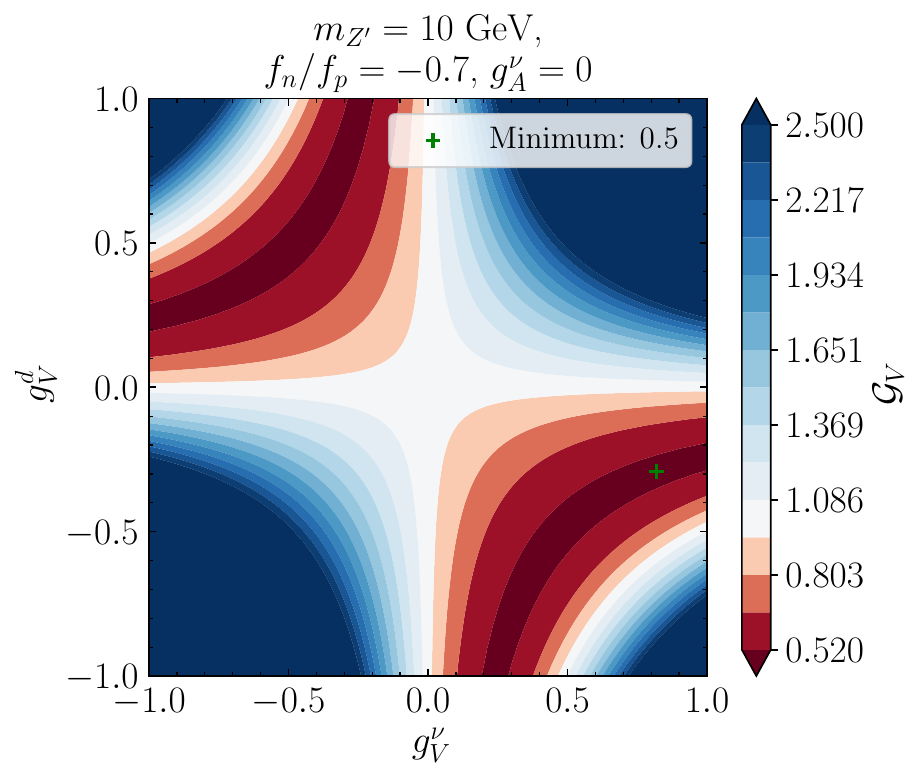}\hspace{0.2cm}
        \includegraphics[width=.48\textwidth]{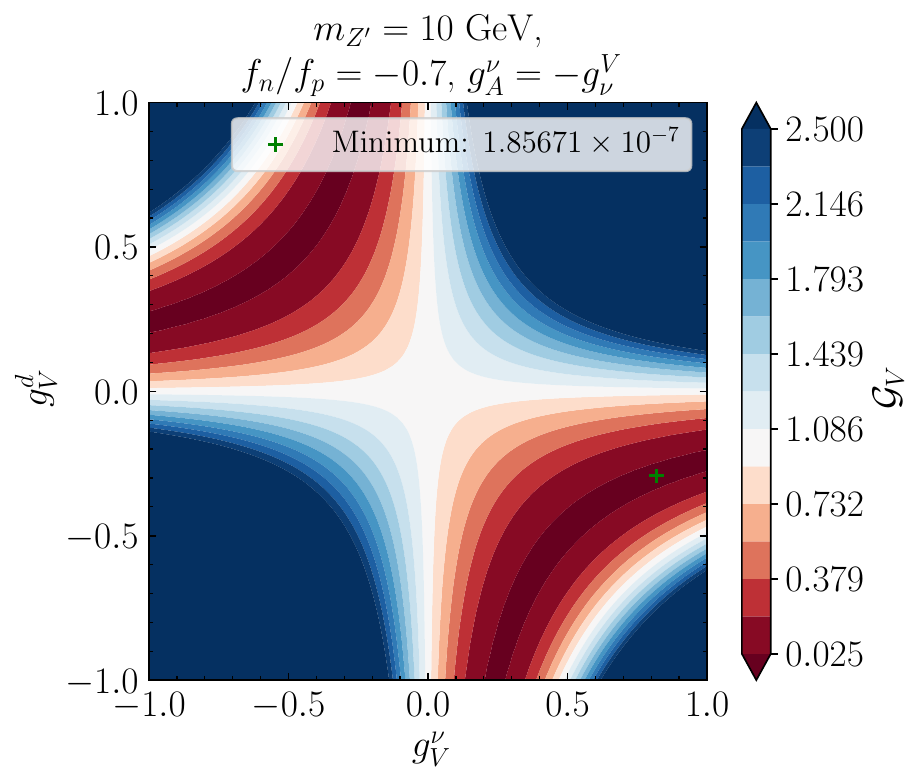}\\
        \includegraphics[width=.48\textwidth]{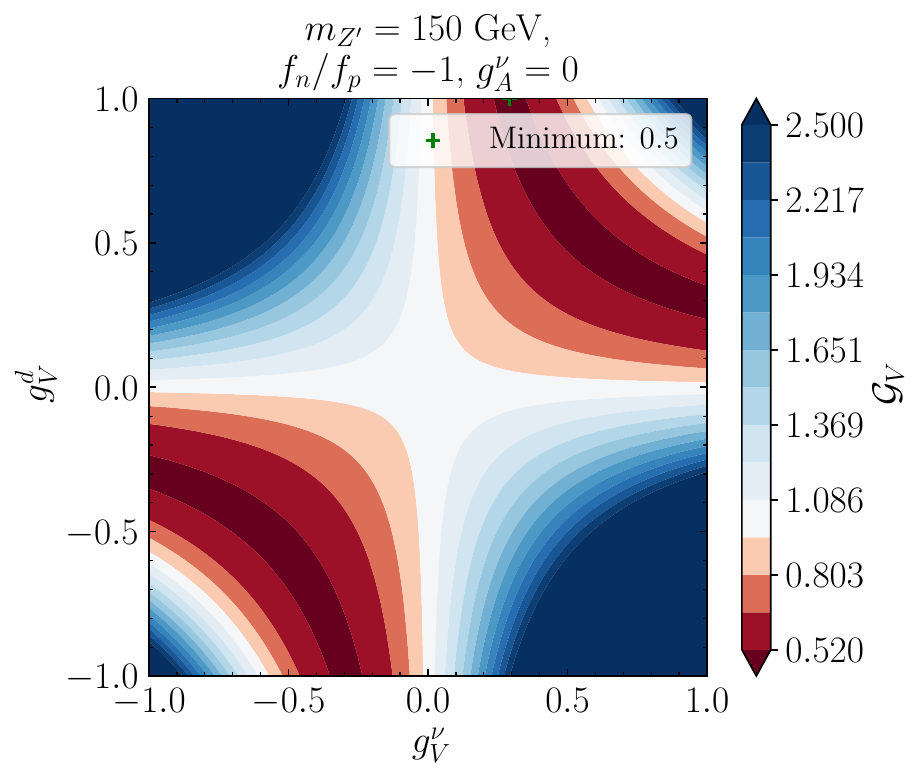}\hspace{0.2cm}
        \includegraphics[width=.48\textwidth]{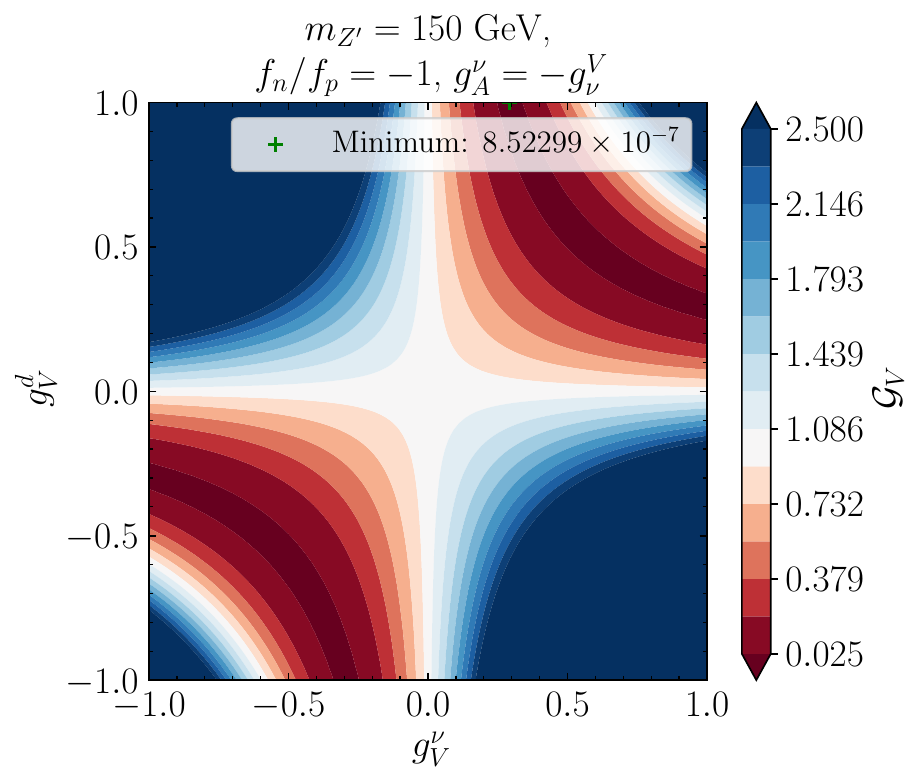}
	\caption{Contour plot of \GV\ as a function of $g^{\nu}_V$ and $g_V^d$, for $f_n/f_p=-0.7$ and $m_{Z'}=10$ GeV (top row), and $f_n/f_p=-1.0$ and $m_{Z'}=150$ GeV (bottom row), and for $g_A^\nu=0$ (left column) and $g_A^\nu=-g_V^\nu$ (right column). Red-scale contours correspond to $\mathcal{G}_V<1$, and blue-scale ones to $\mathcal{G}_V>1$, while the white areas correspond to $\mathcal{G}_V$ values close to $1$. The minimum possible value of $\mathcal{G}_V$ is shown in the top right corner of each subfigure, with its position in the plot marked as a green cross. All plots were computed at zero recoil energy $E_R$ (note from Eq. \eqref{eq:GV} that typical values of $E_R$ will produce negligible changes in $\mathcal{G}_V$).}
	\label{fig:gv2a}
\end{figure*}

As it can be seen in Eq.~\eqref{eq:GV}, the interference is driven by the neutrino couplings to the $Z'$, $g_{V,A}^\nu$, and the sign of $Q_V$, which depends on the values of $g_{V}^{u,d}$ and $f_n/f_p$. In Fig.~\ref{fig:gv2a} we show the resulting values of \GV\ as a function of $g_V^d$ and $g_V^\nu$, for different values of $f_n/f_p$ and $g_A^\nu$. On the top row of Fig.~\ref{fig:gv2a} we can see the values of \GV\ for $f_n/f_p=-0.7$ and a $Z'$ mass $m_{Z'}=10$ GeV. On the left panel we present the case where $g_A^\nu=0$, so the neutrino only has vector interactions with the $Z'$. In that case the factor \GV\ takes values that are smaller than 1 for both positive and negative values of $g_V^d$ and $g_V^\nu$. Therefore, the minimum value of \GV\ that can be reached is $\mathcal{G}_V=0.5$. In the cases where $g_V^d$ and $g_V^\nu$ have opposite signs we found that $\mathcal{G}_V>1$. On the right panel the same case is depicted but taking $g_A^\nu=g_V^\nu$. The effect of taking this value can be seen as a reduction of the minimum value of \GV, that in this case can go down to $\mathcal{G}_V=1.46\times 10^{-6}$. A similar behavior is found for other configurations of $f_n/f_p$. In the lower row of Fig.~\ref{fig:gv2a} we show the results for $f_n/f_p=-1$ and a $Z'$ mass of $m_{Z'}=150$ GeV. Left and right panels have $g_A^\nu=0$ and $g_A^\nu=g_V^\nu$ respectively as in the previous case. We observe that, for the same sign of $g_V^d$ and $g_V^\nu$, the value of \GV\ is greater than one, while smaller values, $\mathcal{G}_V<1$ are achieved when the two couplings have different signs. 

\subsubsection{Number of Events}

As we mentioned before, the presence of \GV\ in the \cevns\ differential cross section will impact the number of neutrino events hitting the detector. The differential number of events per unit energy produced by neutrinos can be written as
\begin{equation}
    \frac{dN_{\nu-N}}{dE_R}=\frac{\varepsilon}{m_N}\int_{E_\nu^{\rm min}}dE_{\nu} \mathcal{A}(E_R) \frac{d\Phi}{dE_\nu}\frac{d\sigma^\nu}{dE_R},
    \label{eq:nuevent}
\end{equation}
where $\varepsilon$ is the exposure of the detector, $m_N$ is the target nuclear mass, $d\Phi/dE_\nu$ is the incoming neutrino flux, and $\mathcal{A}(E_R)$ is the detector efficiency that we set to one for simplicity. Depending on the material of the detector, the recoil of the nucleus will take place only for energies above a certain energy threshold of the incoming neutrinos, $E_\nu^{\rm min}$, that can be written as $E_\nu^{\rm min}=\sqrt{m_NE_R/2}$ for $m_N \gg E_R$. 

\begin{figure*}[ht]
	\centering
        \includegraphics[width=.44\textwidth]{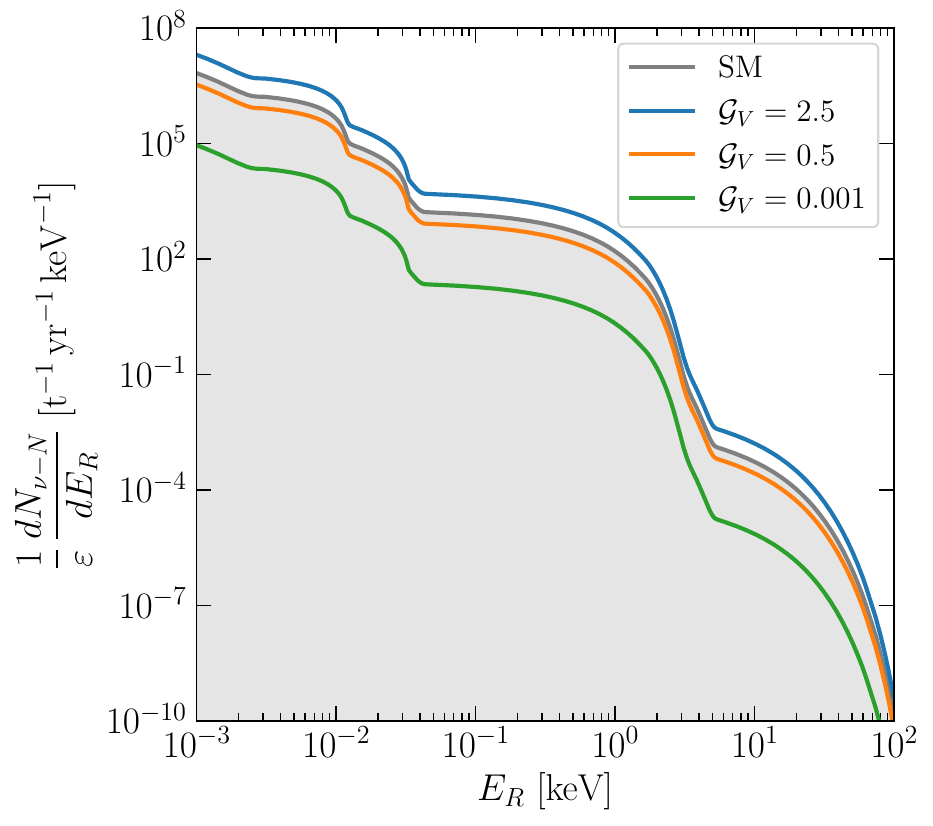}\hspace{1.0cm}
        \includegraphics[width=.44\textwidth]{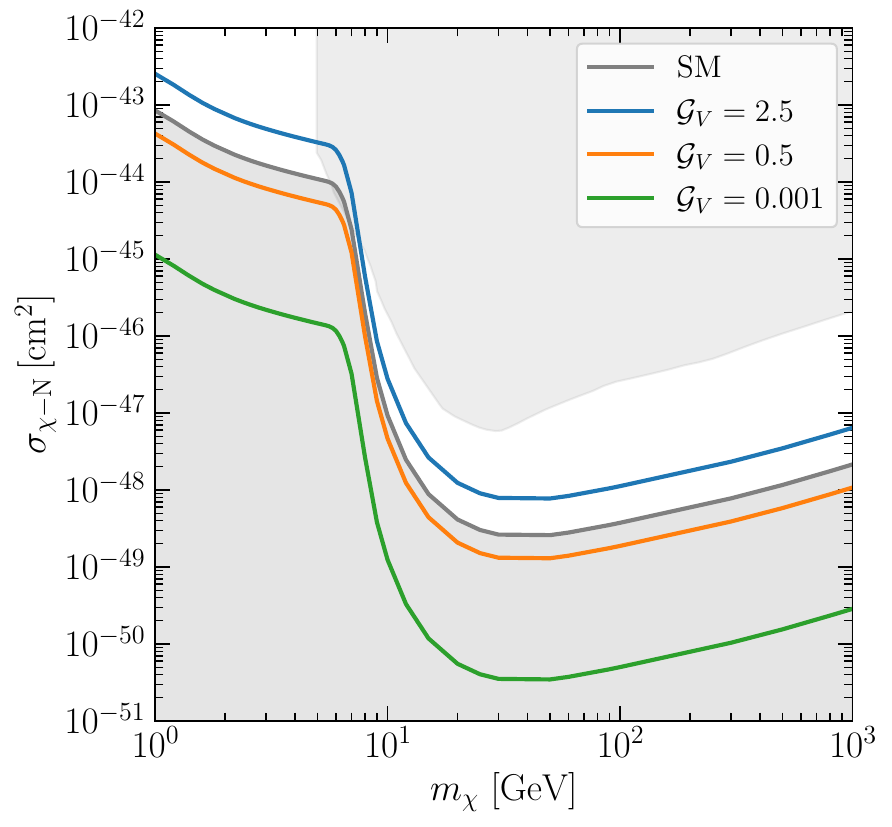}
	\caption{\textbf{Left Panel:} Differential number of neutrino events per ton and per year of exposure in a Xenon detector as a function of recoil energy for different values of \GV$=2.5$ (blue), 0.5 (orange) and 0.001 (green). The SM prediction for the number of neutrino events is shown as a gray area. \textbf{Right Panel:} Neutrino floor for Xe-based experiments as a function of DM mass for different values of \GV$=2.5$ (blue), 0.5 (orange) and 0.001 (green). The SM prediction is shown as a gray area for comparison, and isospin-conserving bounds from Xe-based direct detection experiments as a light gray area.}
	\label{fig:gvevents}
\end{figure*}
In order to compute the total number of neutrino events, one can integrate Eq.~\eqref{eq:nuevent},
\begin{equation}
    N_{\nu-N} = \int_{E_{\text{th}}} dE_R \frac{dN_{\nu-N}}{dE_R}\,,
    \label{eq:nueventrate}
\end{equation}
where $E_{\text{th}}$ denotes the recoil energy threshold. In the left panel of Fig.~\ref{fig:gvevents} we show the differential number of neutrino events per unit recoil energy and per ton and year of exposure of a Xe detector as a function of the recoil energy for different values of \GV. For these numerical computations, we have used the incoming neutrino flux from Ref.~\cite{Strigari:2009bq}. We show three values of \GV, 2.5, 0.5, and 0.01, as well as the SM value for comparison. As we can see, the neutrino event rate profile is similar to the incoming neutrino flux; however, the different values of \GV\ can alter directly the rate.  As we expected from the definition of \GV\ in Eq.~\eqref{eq:xi}, different values will enhance or decrease the expected number of neutrino events in the detector.

Direct detection experiments were designed to detect DM in such a way that the possible background could be inexistent. However, as we mentioned before, the incoming neutrino flux can produce a nuclear recoil in the detectors. Once this point is reached, the search for DM is no longer background free. Given this fact, one can compute the minimum cross section for each DM mass from which the incoming neutrinos act like background, namely the neutrino floor~\cite{Strigari:2009bq}. In order to compute it we follow the one-neutrino event approach as in Refs.~\cite{Billard:2013qya, Bertuzzo:2017tuf, Boehm:2018sux, Sadhukhan:2020etu}.\footnote{An statistical approach taking into account statistical significances for the neutrino detection can be found in, for instance, Refs.~\cite{Bertuzzo:2017tuf, AristizabalSierra:2021kht}.} This approach consists of considering the exposure required to have 1 neutrino event produced by \cevns\ in the detector for a given target nucleus with a minimum energy threshold, $E_{\rm th}^{\rm min}$. According to Eq.~\eqref{eq:nueventrate} this can be written as
\begin{equation}
    \varepsilon (E_{\rm th}) = \frac{1}{\int_{E_{\rm th}} dE_R \frac{dN^\nu}{dE_R}},
    \label{eq:1nevent}
\end{equation}
where we have factorized $\varepsilon$ from Eq.~\eqref{eq:nuevent}. This exposure is to be calculated for every $E_{\rm th}$ in order to obtain the background-free exclusion limit, which at 90\% C.L. corresponds to 2.3 DM events.

The differential DM-nucleus number of events can be calculated as
\begin{equation}
    \frac{dN_{\chi-N}}{dE_R}=\varepsilon \frac{\rho_{\rm DM} \sigma_0^A}{2 m_\chi \mu_n^2}\mathcal{F}^2(E_R)\int_{v_{\rm min}}d^3v\frac{f(v)}{v},
\end{equation}
where $\varepsilon$ is the exposure, $\rho_{\rm DM}=0.3$ GeV$/{\rm cm}^3$ is the total DM density, $\sigma_0^A$ is the cross section defined in Eq.~\eqref{eq:sigma0}, $\mathcal{F}$ is the Helm form factor, $m_\chi$ is the DM mass, and $\mu_n$ is the reduced nucleon-DM mass. $f(v)$ is the DM velocity distribution according to the reference frame of the Earth, and we will assume its profile as a Maxwell-Boltzmann distribution. The integral that has a lower limit $v_{\rm min}=\sqrt{m_NE_R/2\mu_N^2}$, where $\mu_N$ is the nucleus-DM reduced mass, can be computed analytically, obtaining
\begin{equation}
    \int_{v_{\rm min}}d^3v\frac{f(v)}{v}=\frac{1}{2v_0\eta_E}[{\rm erf}(\eta_+) - {\rm erf}(\eta_{-})] - \frac{1}{\pi v_0 \eta_E}(\eta_+ - \eta_-)e^{-\eta^2_{\rm esc}}\,.
\end{equation}
Here $\eta_E=v_E/v_0$, with $v_E=232$ km/s, the velocity of Earth with respect to the galactic center, and $v_0=220$ km/s, the local galactic rotational velocity, $\eta_{\pm}={\rm min}(v_{\rm min}/v_0\pm \eta_E, v_{\rm esc}/v_0)$, with $v_{\rm esc}=544$ km/s the escape velocity of DM from our galaxy. In a similar fashion to Eq.~\eqref{eq:nueventrate}, one can define the number of events from the differential number of events with
\begin{equation}
    N_{\chi-N} =\int_{E_{\rm th}}dE_R\frac{dN_{\chi-N}}{dE_R}\,.
\end{equation}
The procedure now consists of computing the DM number of events equal to 2.3 as in Eq.~\eqref{eq:1nevent} obtaining the next relation,
\begin{equation}
    \int_{E_{\rm th}}dE_R\frac{dR^\chi}{dE_R}=\frac{2.3}{1}\int_{E_{\rm th}}dE_R\frac{dN^\nu}{dE_R}\,.
\end{equation}
From this relation, a number of DM-nucleon scattering cross section isocurves as a function of the DM mass are obtained for varying threshold energies, $E_{th}$, in logarithmic steps from 0.001 keV to 100 keV. From those isocurves generated with the different values of $E_{th}$, we take the lowest cross section for each mass to obtain the limit at 90\% C.L. in which the DM detection searches are background free.

As we have mentioned before, the presence of new interactions in the neutrino sector will affect the expected number of events hitting the detectors. This certainly will affect the neutrino floor since the interactions between the nucleus and neutrinos get modified with respect to the SM. In the right panel of Fig.~\ref{fig:gvevents} we can see the effect of different \GV \ factors on the neutrino floor for a xenon-based detector. 
We have used as an example the following values for $\mathcal{G}_V$: 2.5 (blue), 0.5 (orange) and 0.001 (green), adding the SM prediction as a gray area for comparison. Different values of the factors will change the value of the neutrino floor with respect to the SM. If the parameter $\mathcal{G}_V>1$, the neutrino floor will present a higher cross section, while for $\mathcal{G}_V<1$ the cross section will present lower values with respect to that of the SM. This modification can be constrained in the future if observations of neutrino events are made in direct detection experiments.

\section{Results}
\label{sec:results}

 In this section we will show the effects of isospin-violating interactions in both the direct detection bounds and the neutrino floor. In order to cover the different phenomenology that one can find in this kind of models, we categorize scenarios based on their behaviour with respect to the parameter space left between the neutrino floor and the experimental DM bound, relative to the SM prediction, represented qualitatively by the quantity $\Delta$. If $\Delta=0$, the parameter space left between the direct detection dark matter bounds is unchanged from the SM prediction; $\Delta>0$ indicates that the distance between the direct detection bounds and the neutrino floor opens up, and $\Delta<0$ means this gap will be smaller than in the SM. In Table~\ref{tab:scenarios_adrian} we present different combinations giving rise to the scenarios considered in this work, based on the changes of the dark matter bounds and the neutrino floor with respect to the isospin-conserving case. The scenarios considered in this work are classified as follows:

 \begin{itemize}
    \item {\bf Type I ($\Delta<0$)}: In this case ({\bf Scenario I}), the effects of isospin violating interactions do not affect the dark matter bounds, but make the neutrino floor rise up, hence reducing the parameter space and inducing more overlap between the neutrino floor and the experimental DM bound than in the isospin-conserving case.
    \item {\bf Type II ($\Delta=0$)}: Here, the parameter space remains the same as in the isospin-conserving case. This can be achieved in two scenarios:
    \begin{itemize}
        \item {\bf Scenario IIa}: The neutrino floor and the direct detection bound are lowered by approximately the same extent.
        \item {\bf Scenario IIb}: The neutrino floor and the direct detection bound are raised by approximately the same extent.
    \end{itemize}
    Even if both situations mimic the isospin-conserving scenario, the observation of the neutrino floor in these cases would happen in lower or higher DM-nucleon scattering cross sections.
    \item {\bf Type III ($\Delta>0$)}: The parameter space in this case is opened up with respect to the isospin-conserving case. The scenarios of this type are ordered with respect to the amount of parameter space opened up:
    \begin{itemize}
        \item {\bf Scenario IIIa}: both floor and bound are lowered, but the floor goes lower.
        \item {\bf Scenario IIIb}: both floor and bound are raised, but the bound goes higher.
        \item {\bf Scenario IIIc}: the floor is lowered and the bound remains unchanged.
        \item {\bf Scenario IIId}: the floor is lowered and the bound is raised, opening up a lot of parameter space with respect to the SM.
    \end{itemize}
\end{itemize}

\begin{table}[ht]
\centering
\begin{tabular}{ccccc|}
&   & \multicolumn{3}{c}{Neutrino floor}   \\
\hhline{~~|-|-|-|}
& \multicolumn{1}{c|}{}  & \multicolumn{1}{c|}{\parbox[c]{2cm}{\centering Down}} & \multicolumn{1}{c|}{\parbox[c]{2cm}{\centering Unchanged}}    & \parbox[c]{2cm}{\centering Up}    \\
\hhline{~|-|-|-|-|}
\multicolumn{1}{c|}{}   & \multicolumn{1}{c|}{\rotatebox[origin=c]{90}{\parbox[c|]{2cm}{\centering Down}}}    & \multicolumn{1}{c|}{\parbox[c]{2cm}{\centering IIa ($\Delta=0$) \\ IIIa ($\Delta>0$)}}   & \multicolumn{1}{c|}{\cellcolor{gray}} & \multicolumn{1}{c|}{\cellcolor{Brown3} \parbox[c]{2cm}{\color{black} Disfavoured by exp. results}} \\
\hhline{~|-|-|-|-|}
\multicolumn{1}{c|}{\multirow[c|]{3}{*}[1.8cm]{\rotatebox[origin=c]{90}{\parbox[c]{4cm}{\centering Dark matter bound}}}}  & \multicolumn{1}{c|}{\rotatebox[origin=c]{90}{\parbox[c]{2cm}{\centering Unchanged}}} & \multicolumn{1}{c|}{IIIc} & \multicolumn{1}{c|}{\cellcolor{black} {\color{white} SM}} & \multicolumn{1}{c|}{I} \\
\hhline{~|-|-|-|-|}
\multicolumn{1}{c|}{}  & \multicolumn{1}{c|}{\rotatebox[origin=c]{90}{\parbox[c]{2cm}{\centering Up}}}    & \multicolumn{1}{c|}{IIId} & \multicolumn{1}{c|}{\cellcolor{gray}} & \parbox[c]{2cm}{\centering IIb ($\Delta=0$) \\ IIIb ($\Delta>0$)}  \\
\hhline{~|-|-|-|-|}
\end{tabular}
\caption{Pictorical representation of the scenarios considered in this work, based on the change of the neutrino floor and direct detection bounds. $\Delta$ represents the change in the parameter space left between the neutrino floor and the experimental DM bounds with respect to the SM.}
\label{tab:scenarios_adrian}
\end{table}

As in scenario type I, the scenario of type III where the neutrino floor is unchanged is not considered in this work. In Table~\ref{tab:scenarios} we present the different benchmark points chosen to illustrate the different scenarios proposed. In this Table we name the different couplings of the dark sector as well as the amount of isospin violation, $f_n/f_p$ and the changes in the direct detection bounds, $F_Z^{\rm Xe}$, and the neutrino floor, \GV.

\begingroup

\renewcommand{\arraystretch}{1.5} 
\begin{table}[ht]
\centering
\begin{tabular}{|c|ccccccc|}
\hline
    & ~$f_n/f_p$~ & ~$F_Z^{\rm Xe}$~ & ~\GV~ & ~$g_V^\nu$~ & ~$g_A^\nu$~  & ~$g_V^d$~ & ~$m_{Z^{'}}$~[GeV]~
    \\ \hline
    Scenario I & -2.4 & 0.99 & 1.23 & 0.55 & -0.55 & -0.5 & 750
    \\ \hline
    Scenario IIA & 1.7 & 0.50 &  0.52 & 0.6 & -0.6 &  0.6  & 1000
    \\ \hline
    Scenario IIB & 0.4 & 2.40 &  2.38 & 0.5 & 0.0 & 0.04  & 300
    \\ \hline
    Scenario IIIA & 2.0 & 0.40 & 7$\times 10^{-3}$ & 0.6 & -0.6 & 0.67  & 550
    \\ \hline
    Scenario IIIB & -0.7 & 7640 & 1.23 & 0.5 & 0.0 & 0.1  & 10
    \\ \hline
    Scenario IIIC & 0.95 & 1.06 & 0.12 & 0.25 & -0.25 & 0.3 & 400
    \\ \hline
    Scenario IIID & -1.0 & 31.27 & 0.15 & 0.45 & -0.45 & 0.4 & 150
    \\ \hline
\end{tabular}
\caption{Different benchmark scenarios}
\label{tab:scenarios}
\end{table}
\endgroup

\subsection{\texorpdfstring{Type I ($\Delta<0$)}{Type I (Δ<0)}}

This scenario is characterized by a negligible effect of isospin violation in the dark matter sector but sizable corrections to the neutrino floor, causing it to shift upwards in the parameter space. In the top panel of Fig.~\ref{fig:scenarioIandII} we show the results for the Scenario I benchmark. We have depicted the modified direct detection limits as a blue area, and the modified neutrino floor as an orange area. For comparison, we also show the SM neutrino floor and the isospin-conserving direct detection bounds as dashed lines. In this case, the direct detection limits remain unchanged compared to having isospin-conserving interactions. It is interesting to see that this feature can be reached with $f_n/f_p\neq 1$. Using the xenon abundances from Table~\ref{tab:abundances} and Eq.~\eqref{eq:DZ}, this can be achieved when $f_n/f_p\simeq -2.4$, leading to $F_Z\simeq 1$. In this scenario, the DM parameter space is reduced with respect to the isospin-conserving case. This is due to the fact that the neutrino floor rises above the SM prediction. This would lead to an enhancement of the number of neutrino events in direct detection experiments; however, this increase is not too significant with respect to the SM case.

\subsection{\texorpdfstring{Type II ($\Delta=0$)}{Type I (Δ=0)}}

\begin{figure}[ht]
	\centering
        \includegraphics[width=.48\textwidth]{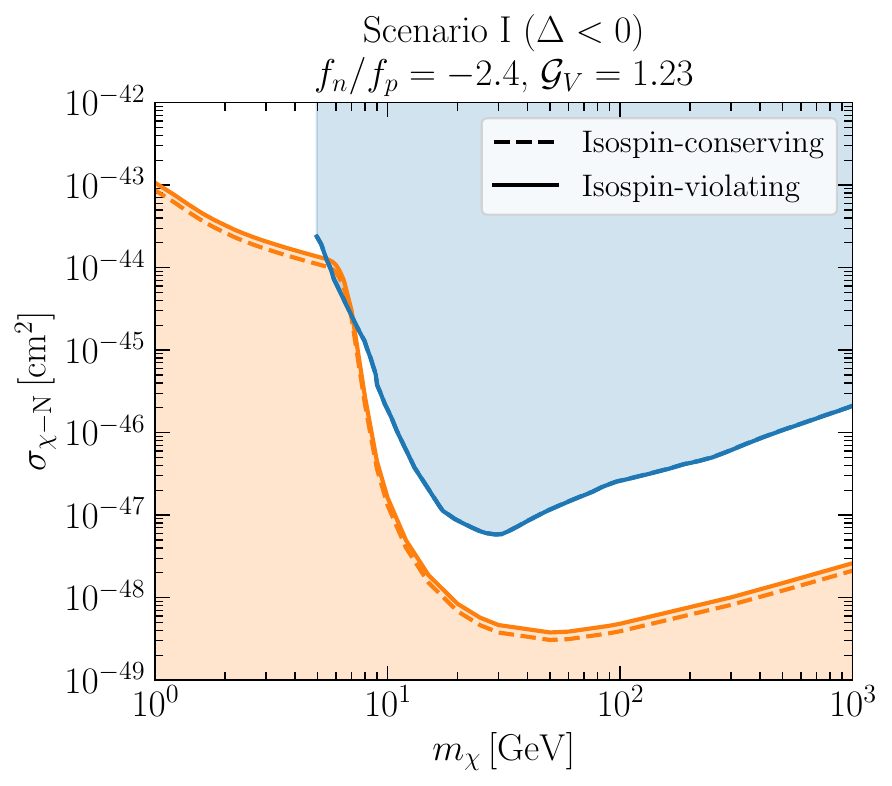}\\
        \includegraphics[width=.48\textwidth]{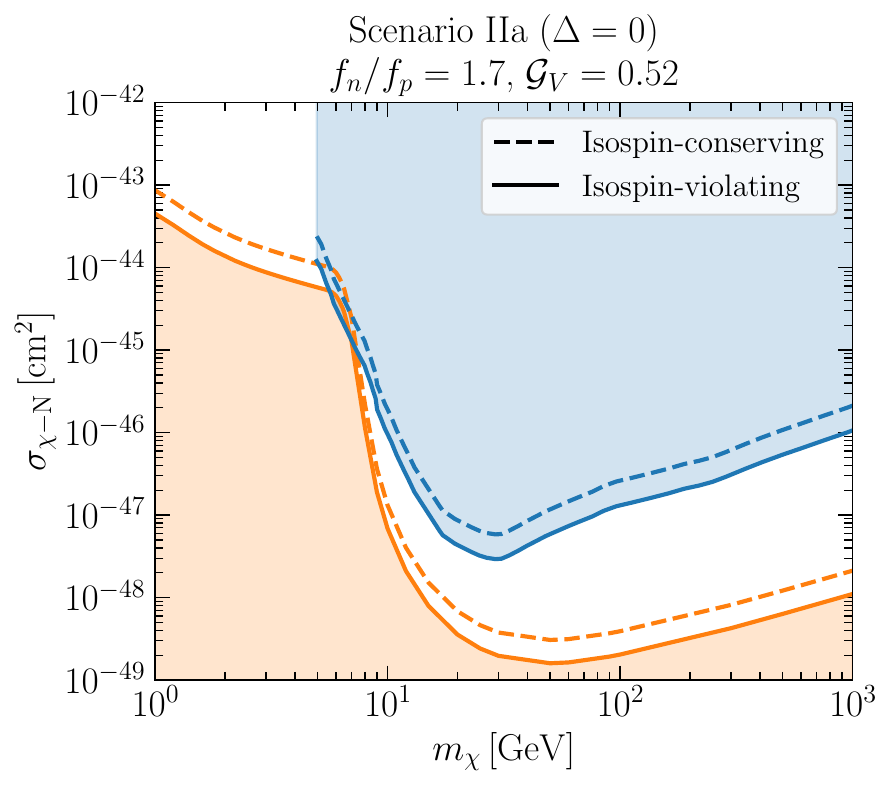}\hspace{0.5cm}
        \includegraphics[width=.48\textwidth]{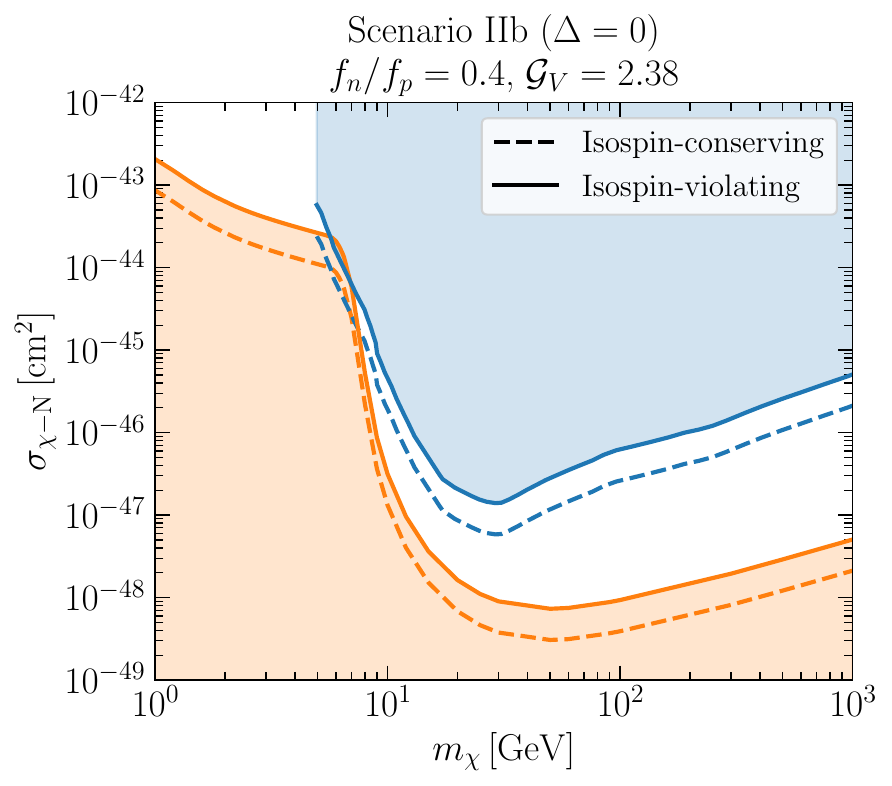}
	\caption{\textbf{Top row:} Direct detection bounds and neutrino floor for Scenario I (with $\Delta < 0 $) \textbf{Bottom row:} Direct detection bounds and neutrino floor for Scenarios IIA and IIB (with $\Delta =0$).}
	\label{fig:scenarioIandII}
\end{figure}

This type of scenario presents a resemblance to the SM case. It is achieved when the direct detection limits and the neutrino floor get shifted by the same amount ($\mathcal{G}_V\simeq F_Z^{\rm Xe}$). On the bottom left panel of Fig.~\ref{fig:scenarioIandII} we can see the results for the Scenario IIA benchmark. In this case, we have $\mathcal{G}_V=0.52$ and the amount of isospin violation is given by $f_n/f_p=1.7$, which leads to a modification of the direct detection bounds of $F_Z^{\rm Xe}=0.5$. The neutrino floor gets reduced relative to the SM one due to the interference terms appearing in Eq.~\eqref{eq:GV}. Furthermore, the amount of isospin violation, $f_n/f_p=1.7$, enhances the direct detection cross section of DM, $\sigma^{\rm SI}$, which consequently leads to more stringent bounds. However, from the point of view of direct detection experiments, the situation would be the same as with isospin-conserving interactions and SM neutrino events. In Scenario IIB (bottom right panel of Fig.~\ref{fig:scenarioIandII}) the situation is the opposite. Here the neutrino floor is enhanced, with $\mathcal{G}_V= 2.38$ , while the amount of isospin violation, $f_n/f_p=0.4$, reduces the dark matter direct detection cross section bound, with $F_Z^{\rm Xe}=2.4$. In both cases the parameter space left for DM detection is the same as in the usual scenario with isospin-conserving interactions and the SM neutrino floor.


\subsection{\texorpdfstring{Type III ($\Delta>0$)}{Type I (Δ>0)}}

\begin{figure*}[ht]
	\centering
        \includegraphics[width=.48\textwidth]{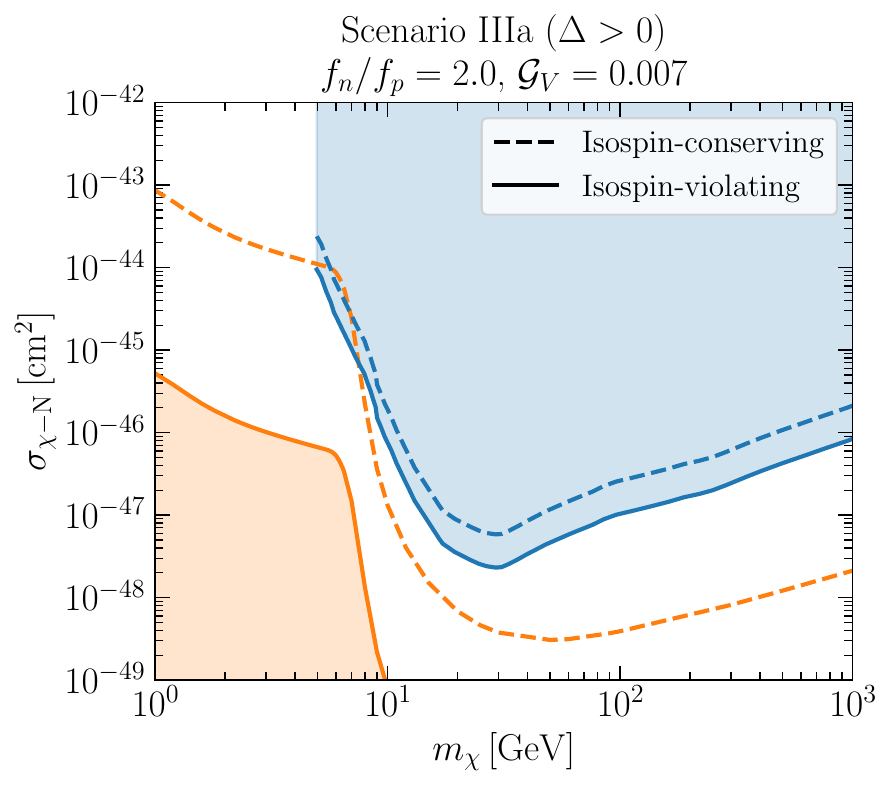}\hspace{0.5cm}
        \includegraphics[width=.48\textwidth]{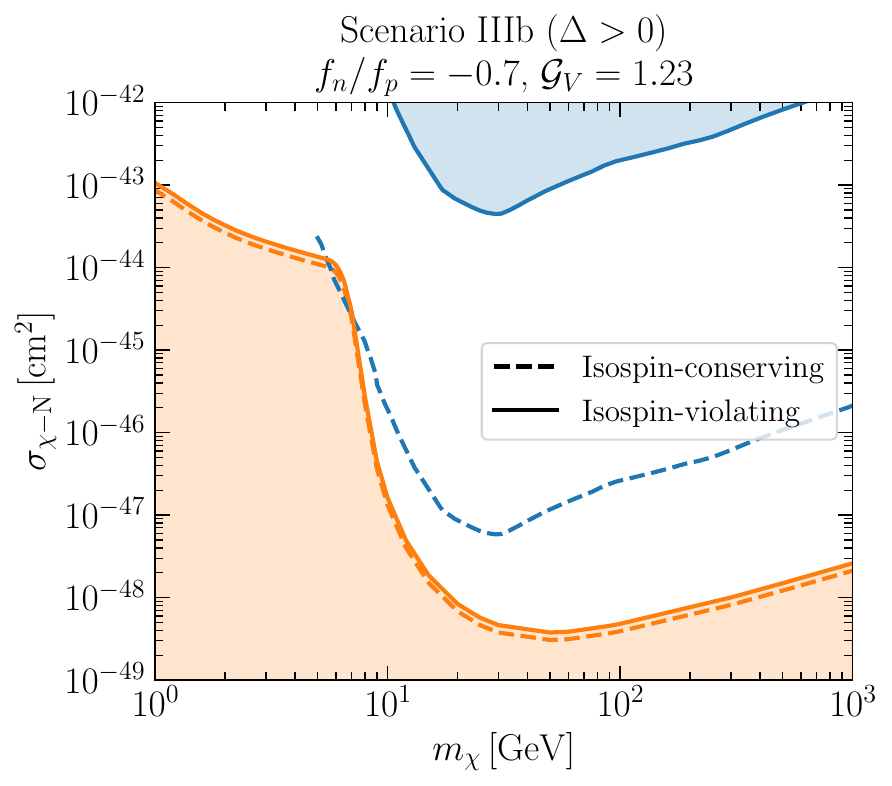}
        \includegraphics[width=.48\textwidth]{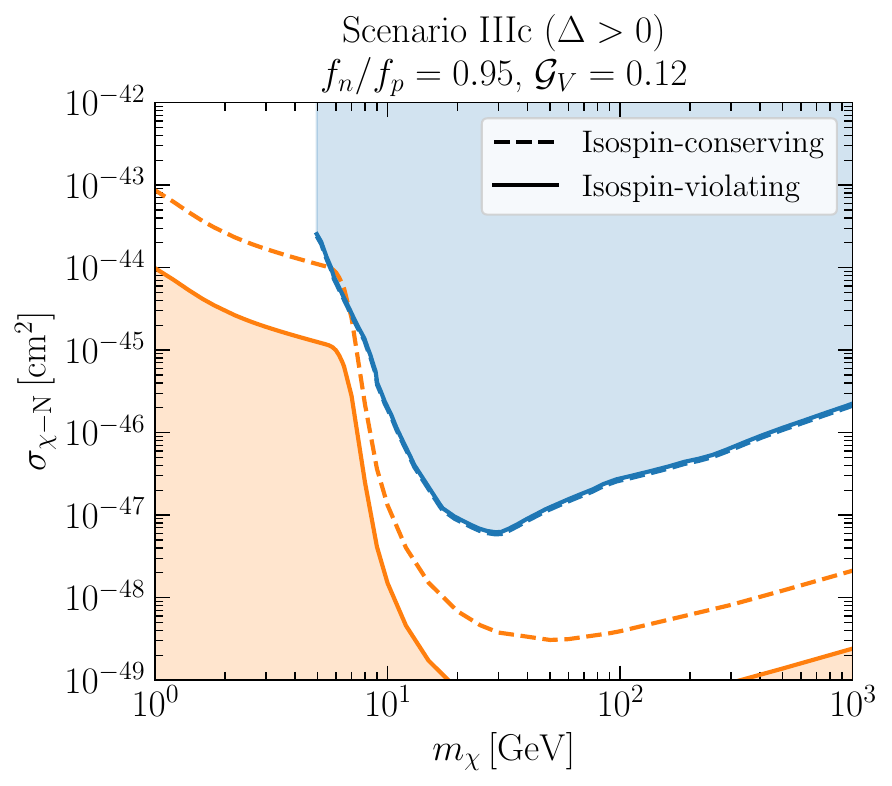}\hspace{0.5cm}
        \includegraphics[width=.48\textwidth]{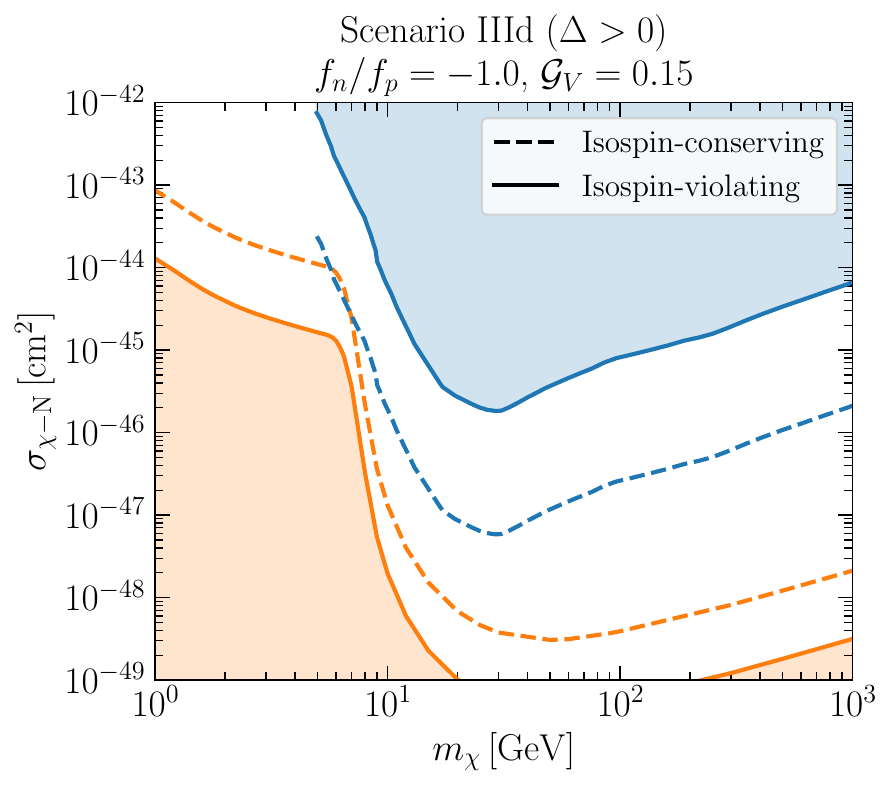}
	\caption{Direct detection bounds and neutrino floor for Scenarios IIIA, IIIB, IIIC, IIID (with $\Delta > 0$).}
	\label{fig:scenarioIII}
\end{figure*}

The last scenarios are defined by a parameter space with a larger separation between the DM direct detection lower bound and the neutrino floor, compared to isospin-conserving interactions. The results of the benchmarks chosen for type III scenarios are shown in Fig. \ref{fig:scenarioIII}. This opening up of the parameter space can be slightly achieved by lowering both DM bound and neutrino floor, but more so the floor (Scenario IIIa, top left panel, with $F_Z^{\rm Xe}=0.4$ and $\mathcal{G}_V=7.0\times 10^{-3}$); or by raising both but more so the DM bound (Scenario IIIb, top right panel, with $F_Z^{\rm Xe}=7640$ and $\mathcal{G}_V=1.23$). Parameter space can also be opened up by a negligible change in the DM sector by the isospin-violating interactions but a sizable one in the neutrino floor, such that it is lowered. This can be achieved with couplings leading to a $f_n/f_p$ value close to 1 (hence a $F_Z^{\rm Xe}$ also close to 1), but such that $\mathcal{G}_V$ is small, e.g. $\mathcal{G}_V = 0.114$ in our benchmark for Scenario IIIC (bottom left panel). Lastly, the most obvious way to separate DM bound and neutrino floor is to shift upwards the bound and downwards the floor, as done in Scenario IIID (bottom right panel), with $F_Z^{\rm Xe}=31.27$ and $\mathcal{G}_V=0.15$.


\newpage

\subsection{Potential Observation of Solar Neutrinos} 

The construction of the neutrino floor involves astrophysical neutrino sources, with solar neutrinos being the most dominant among them. 
In dark matter direct detection experiments, to have a 90\% confidence level for a particular nuclear recoil to be a dark matter signal, it is expected to observe excess events than the predicted neutrino background events: 2.3 dark matter events per neutrino event. 
No direct detection experiment has yet confirmed the observation of a DM event. 
If the projected or observed exclusion limit of any DM direct detection experiment goes below the neutrino floor, then the DM detection becomes implausible for that case as the available DM-nucleon scattering cross section that can be probed can only produce fewer recoil events required to confirm a DM event. In this scenario, either the neutrino floor should be modified to go below the experimental limit or the DM detectors start probing directly the recoils coming from the background neutrinos. 

 In the absence of new parameter space opening up to be looked into for the dark matter discovery, 
direct detection experiments directly probe the neutrino background \textit{i.e.}, the solar neutrinos. 
Very recently, hints of solar neutrino events have appeared in the PandaX-II~\cite{PandaX:2024muv} and Xenon-nT~\cite{XENON:2024ijk} experiments. 
Even in the absence of BSM physics this observation was expected, as the experimental upper bound on the DM-nucleon recoil cross section for the DM detection was touching or crossing the theoretical lower bound coming from the neutrino background in the SM. 
In this case, the allowed DM-nucleon cross section values are not sufficiently enhanced compared to the corresponding background neutrino-nucleon CE$\nu$NS scattering cross section, failing to determine any recoil event to be solely from the DM. 
That means that the detector is only capable of observing any cross section as high as the experimental limit or lower. 
Therefore, DM direct detection experiments can now be used to observe the background neutrinos through their nuclear recoil. 
If the observations of these two DM direct detection experiments confirm their detection with certainty, then a dedicated analysis is required to point out their exact nature and potential ramifications for the DM detection, which we leave for later. 

\begin{figure*}[ht]
	\centering
     \includegraphics[width=.48\textwidth]{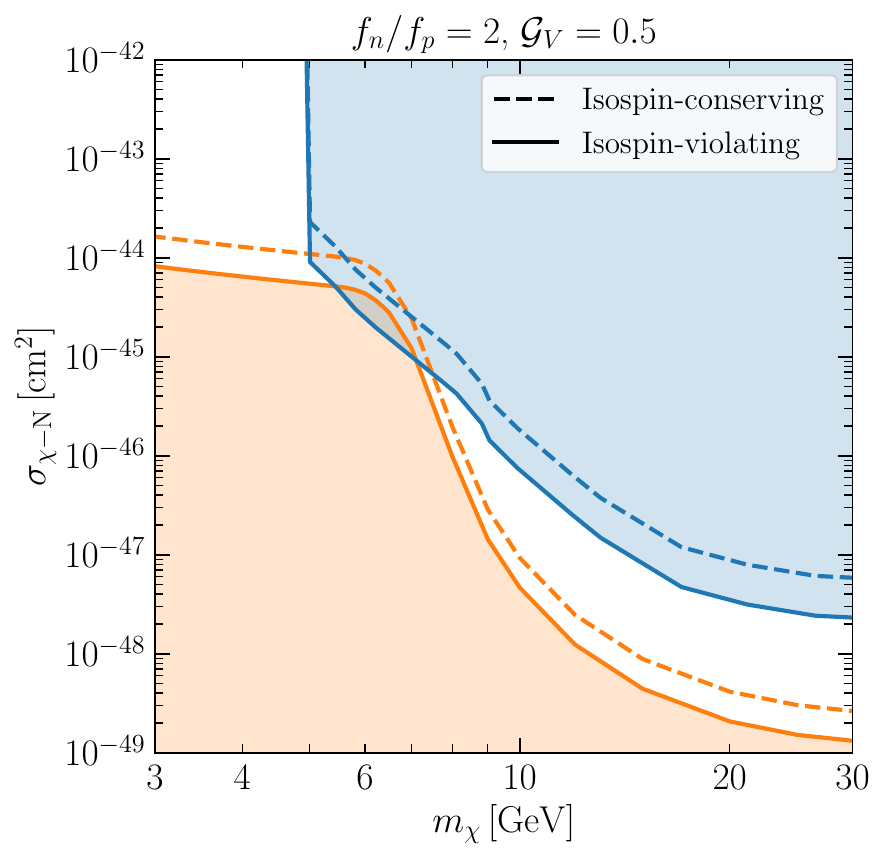}\hspace{0.5cm}
     \includegraphics[width=.48\textwidth]{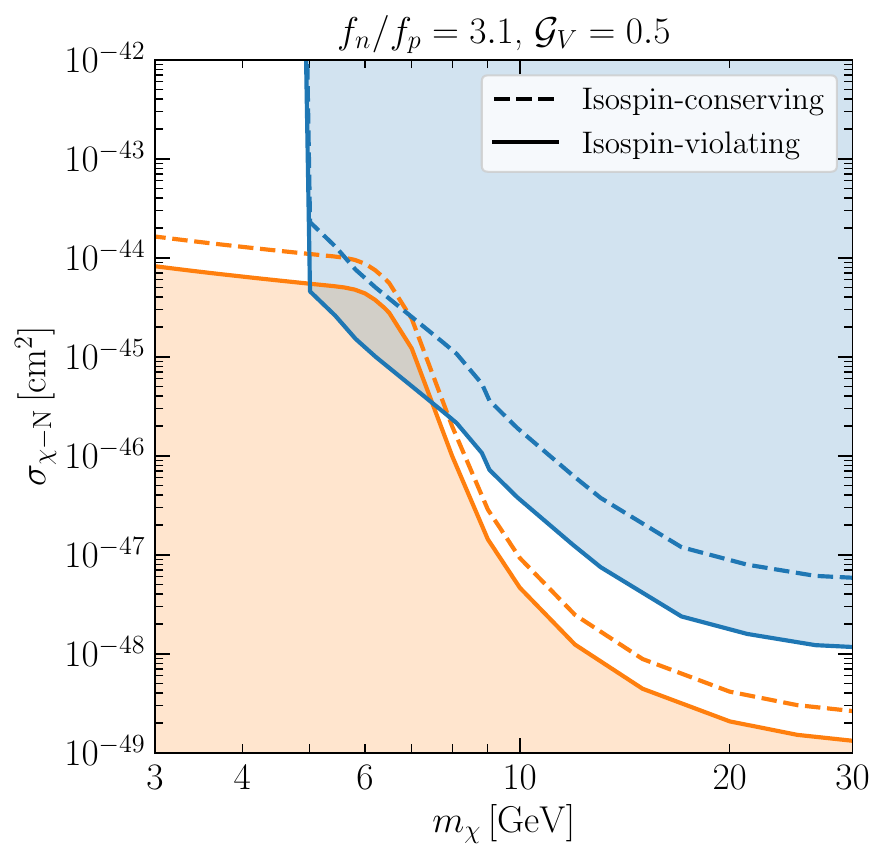} \\
     \includegraphics[width=.48\textwidth]{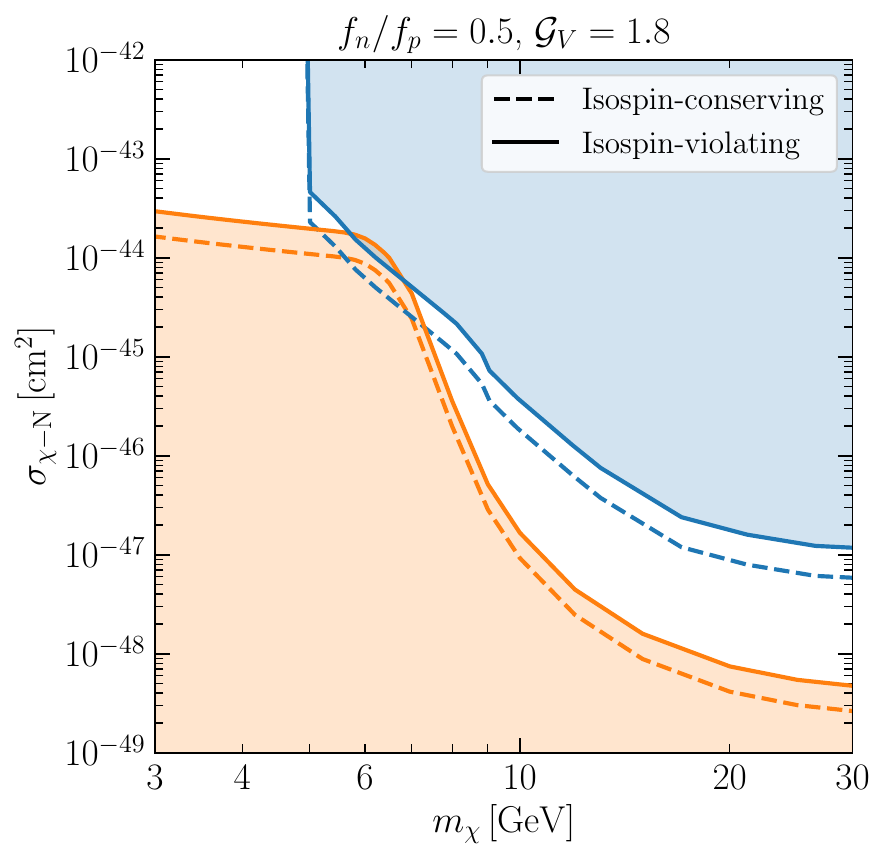}
     \caption{Experimental upper bound in xenon-based dark matter direct detection experiments touching the neutrino floor, computed in the isospin violating model, compatible with the possible observation of neutrino events.}
	\label{fig:gv2b}
\end{figure*}

In this work we analyze the observation potential of solar neutrinos in Xenon-based dark matter direct detection experiments.

Xenon-based experiments have already put the strongest upper bound on allowed values of the DM-nucleon scattering cross section for a wide range of DM masses. 
Future projections of these experiments appears to be tighter, i.e. the allowed DM-nucleon cross section will plunge to lower values. 
If the SM neutrino floor is confronted with the future results, for a significant range of the DM masses, the allowed DM-nucleon cross section is smaller than the one predicted from the SM theory. 
This brings us to the classic dilemma that if the experimental results do not observe the solar neutrinos responsible for the neutrino floor, then the SM neutrino-nucleon interaction rate must be modified to come to a lower value. 
On the other hand, even when we observe solar neutrinos in future experiments, the actual neutrino-nucleon cross section in the SM must be smaller than the computed one, because the experiment is only sensitive to a lower DM-nucleon cross section. 
On conservative side, we propose a condition for the SM neutrino observation, where the experimental result is expected to exactly touch the neutrino floor. 
Only then no parameter space will be left for the DM discovery and along with that the solar neutrinos, if observed, will be seen with effective DM-nucleon cross section for its discovery, corresponding to the neutrino nucleon scattering cross section computed in our model.

In this work the expected future bounds from the experiments are allowing only smaller cross sections of the DM-nucleon scattering than the corresponding strength of the neutrino-nucleon interaction. 
The model that is explored here has isospin-violating dark sector which includes extra $Z^{\prime}$ mediated interaction in both the neutrino and DM sector. 
Presence of isospin violation impacts in two different sectors: one in the scaling of the DM direct detection experimental results and also in the modification of the neutrino floor. 
In general, the non-observation of the DM signal is translated into upper bounds on the allowed DM interaction cross section assuming the SM like interaction with no isospin violation. 
Due to presence of isospin violation in the dark sector, the experimental direct detection result will either scale up or scale down depending on the amount of isospin violation parametrized by the factor $f_n/f_p$. 
On the other hand, CE$\nu$NS gets modified due to isospin-violating coupling of $Z^{\prime}$ with the neutrinos, which is quantitatively encoded in the parameter $\mathcal G_V$. 
We plot different scenarios with different $f_n/f_p$ and $\mathcal G_V$ values. 
In top left plot in Figure.~\ref{fig:gv2b}, $\mathcal G_V$ is taken as 0.5 which indicates that the neutrino floor will come down. 
The isospin-violating parameter $f_n/f_p$ is taken equal to 2 which eases the experimental upper bound allowing for higher cross sections to be probed. 
While in the SM, the theoretical neutrino background cross section is higher than the corresponding experimental limit, the isospin violation can match both of them at DM mass around $m_{\rm DM} \sim 10 ~$GeV. 
This is indicative of some components of the solar neutrinos to be observed for this type of isospin violation. 
In the other plots of Figure.~\ref{fig:gv2b}, similar situations arise where the experimental limit is touching the neutrino floor enabling our limiting case for neutrino detection, albeit preferring DM masses either lower or higher than $10~$GeV. 
For these plots the amount of isospin violation is different, leading to different scaling of the neutrino floor, along with the upper bound in the experiments.  
 
This is the scenario even when presence of any new physics is not taken into account. 
In the SM itself the neutrino floor and leading experimental results from the detection sector are currently coexisting in a way that the neutrino background observation is very much expected. 
Even if the experimental results are not statistically conclusive, we see some hints of the neutrino background observation which can be put as an indication of DM not being present in the mass range we are exploring here.


\section{Conclusions}
\label{sec:conclusions}
In this work, we have explored the direct detection potential of a generic dark sector with isospin violation. 
We focus on a scenario where the gauge sector is extended with a $U(1)_{\rm X}$ symmetry, introducing a new neutral gauge boson, $Z^{\prime}$. 
This extra $Z^{\prime}$ mediates both the DM-nucleon scattering process along with the neutrino nucleus scattering of CE$\nu$NS type. 
The isospin violating couplings $g_V^{\nu}, g_A^{\nu}, g_V^{d}$ of the gauge boson $Z^{\prime}$ are varied over different benchmark scenarios to obtain different phenomenological aspects related to dark matter detection prospects. 

Along with the gauge couplings mentioned above, dark matter nucleon scattering is modified through two new parameters $f_n$ and $f_p$ which essentially leads to one single parameter taken as the ratio $f_n/f_p$. 
The $f_n/f_p =1$ case is the isospin conserving case and isospin violation is included in this part through varying the ratio from $\sim -2$ to 2. 
The effect of the isospin violation scales up the experimental limit for the negative  $f_n/f_p$. 
This leads to more relaxed upper bound from the DM direct detection experiments, pushing the bound on the DM-nucleon cross section to higher values. 
The relaxation of the experimental limit is maximum for the isospin violation of $f_n/f_p = - 0.7$. 
While the positive  $f_n/f_p$ can relax the experimental bound for small values, but for large positive values the bound gets stricter. 

On the other hand, isospin violation in the neutrino sector modifies CE$\nu$NS and that modifies the neutrino nucleon scattering rate in the DM detectors. 
This modification encoded through the factor $\mathcal G_V$, translates to raising or lowering the neutrino floor which is the effective background cross section for the discovery reach of a DM candidate. 
The neutrino floor which is the lower bound on the DM-nucleon scattering cross section: only for cross sections above it one can differentiate the origin of a nuclear recoil event to be from the DM scattering or neutrino scattering. 
The neutrino floor can go up or go down for different values of the isospin violating parameters in the neutrino sector.

The combination of the effects in the two sectors, one in the rescaling of the experimental result from the DM direct detectors and the another in the modification of the neutrino floor can present different scenarios which can be broadly categorized in the seven scenarios. Different scenarios are quantified through a parameter $\Delta$ which measures the extent of parameter space available for the DM to be discovered. 
This is essentially the region between the experimental upper bound on the allowed DM-nucleon cross section and theoretical lower bound termed as neutrino floor in the $\sigma_{\rm DM-N}- m_{\rm DM}$ plane. 
For the first scenario, the allowed dark matter parameter to be probed further shrinks due to the introduction of the isospin violation, making this case nearly impossible one to look for a dark matter signal. 
There can be two other scenarios where $\Delta =0$, as the neutrino floor and the experimental upper bound both either goes up or goes down simultaneously through a similar extent. 
These cases almost mimic the situation we encounter in the SM, though with the different allowed region of DM-nucleon scattering cross section. 
There are four more cases where extra parameter space for the dark matter discovery opens up due to isospin violation in the dark sector, making them favorable to be probed in the ongoing  and future experiments. 
It can happen either through scaling up of the upper limit of the experimental bounds or through neutrino floor going down or any combined effect of both of these two parameter space relaxing scenarios. 
We finally also discuss the future scope of the DM experiments to look for the effects of isospin violation in the dark sector, even when in the absence of a DM signal, a solar neutrino observation in DM direct detection experiments seems plausible. 
To conclude, as elaborated here, the isospin violating interactions in the DM sector can alter the detection prospects of the DM in different non-trivial manner.

\section*{ACKNOWLEDGMENTS}
We thank Pablo M. Candela and Yuber F. Perez Gonzalez for fruitful discussions about the computations of neutrino interactions. 
The work from V.M.L and A.T. has been supported by the Spanish grants PID2023-147306NB-I00 and Severo Ochoa Excellence grant CEX2023-001292-S (AEI/10.13039/501100011033) and by Prometeo CIPROM/2021/054 from Generalitat Valenciana. 
A.T. is supported by the grant CIACIF/2022/159, funded by Generalitat Valenciana. 
 S.S. thanks Prof. Jose W F Valle for the opportunity of an academic visit to IFIC, Valencia where the authors started the discussion.  S.S. also thanks SERB (currently ANRF), Govt. of India for supporting this work through the TARE grant with number TAR/2023/000288.

\bibliographystyle{JHEP}
\bibliography{refs}

\end{document}